\documentclass{sigplanconf}

\usepackage{sober}
\usepackage{here}
\usepackage[T1]{fontenc}
\usepackage[utf8]{inputenc}
\usepackage[hypertex,ps2pdf]{hyperref}
\usepackage{subfigure}
\usepackage{array}
\usepackage{xcolor}
\usepackage{amsmath}
\usepackage{amsthm}

\newlength{\mynegesp}

\usepackage{macros}

\renewcommand{\lv}{lv}

\providecommand{\marc}[1]{\textcolor{red}{#1}}
\providecommand{\gwen}[1]{\textcolor{green}{#1}}
\renewcommand{\marc}[1]{}
\renewcommand{\gwen}[1]{}

\title{A Type System for the Automatic Distribution of Higher-order
  Synchronous Dataflow Programs}

\authorinfo{Gwena\"el Delaval}
           {INRIA-IRISA\\Rennes\\France}
        {Gwenael.Delaval@inria.fr}
\authorinfo{Alain Girault}
        {INRIA Rh\^one-Alpes\\Grenoble\\France}
        {Alain.Girault@inria.fr}
\authorinfo{Marc Pouzet}
           {LRI, Univ. Paris-Sud\\Orsay\\France}
        {Marc.Pouzet@lri.fr}

\begin{document}

\conferenceinfo{LCTES'08,} {June 12--13, 2008, Tucson, Arizona, USA.}

\CopyrightYear{2008}

\copyrightdata{978-1-60558-104-0/08/06}

\maketitle

\begin{abstract}
  We address the design of distributed systems with synchronous dataflow
  programming languages. As modular design entails handling both architectural
  and functional modularity, our first contribution is to extend an existing
  synchronous dataflow programming language with primitives allowing the
  description of a distributed architecture and the localization of some
  expressions onto some processors. We also present a distributed semantics to
  formalize the distributed execution of synchronous programs. Our second
  contribution is to provide a type system, in order to infer the localization
  of non-annotated values by means of type inference and to ensure, at
  compilation time, the consistency of the distribution. Our third contribution
  is to provide a type-directed projection operation to obtain automatically,
  from a centralized typed program, the local program to be executed by each
  computing resource. The type system as well as the automatic distribution
  mechanism has been fully implemented in the compiler of an existing
  synchronous data-flow programming language.
\end{abstract}

\gwen{Enlevé : la phrase disant ce qu'on obtient techniquement : ``The
  compilation of this language then produces one ad-hoc binary for
  each physical site. The simultaneous execution of these binaries
  results in a parallel reactive system.''}



\section{Motivations}
\label{sec:intro}

Synchronous programming languages~\cite{synchrony02pieee} are frequently used in
the industry for the design of real-time embedded systems. Such languages define
deterministic behaviors and lie on formal semantics, making them suitable for
the design and implementation of safety critical systems. They are used, for
example, in critical domains such as the automotive, avionics, or nuclear
industry.

Most of the systems designed with synchronous languages are centralized systems.
The parallelism expressed in these languages is a \emph{functional} one, whose
purpose is to ease the design process by providing ideal timing and concurrency
constructs to the designers. A synchronous program is then compiled into a
sequential program emulating the parallel execution of the functional parallel
branches. This sequential program is executed on a single computing resource.
Yet, most embedded systems are composed of several computing resources (named
``locations''), for reasons such as performance, dedicated actuators or sensors
drivers, or adaptivity of the locations to the tasks they are assigned to (e.g.,
pure computing tasks vs control tasks). We call this the \emph{execution}
parallelism. This paper addresses the problem of mapping the functional
parallelism onto the execution one, in a modular way. We focus on distributed
systems implemented as networks of deterministic processes communicating with
FIFOs.

Fragments of a distributed system can be designed separately; but in complex and
multifunctional embedded systems, functionalities are frequently independent of
the hardware architecture, implying conflicts between architectural and
functional modularity. Thus, one functionality can use several locations and one
location can be involved in several functionalities. As a result, programming
each location separately compromises the modularity and is error-prone. This
situation occurs within several industrial areas, such as automotive embedded
systems, and software-defined radio~\cite{mitola95}.


Our paper is organized as follows: Section~\ref{sec:context-motivations} gives
an overview of the context, and motivates our method through some examples and
one application. Section~\ref{sec:formalization} presents the semantics
and the formalization of the spatial type system. Section~\ref{sec:distribution}
presents the projection operation, which is our third contribution. Finally,
related work and discussion about the solution will be exposed in
Section~\ref{sec:discussion}.

\section{Overview}
\label{sec:context-motivations}

\subsection{Distribution of Synchronous Dataflow Programs}
\label{sec:distr-synchr-progr-1}

\marc{Enlever classification lg synch : n'amène pas d'info.}

\newcommand{\Nat}{I\!\!N}


Synchronous dataflow languages, such as Lustre, Signal, or Lucid
Synchrone~\cite{synchrony02pieee}, manipulate infinite streams of values as
primitive values: the notation \texttt{1} represents the infinite stream
$1,1,...$, while \Int stands for the type of infinite streams of integers. For
any stream~$x$, we note $x_i$ its $i$th value. In this context, functions
(called nodes hereafter) are stream functions: e.g., $\Int \fun \Int$ is the
type of functions from integer streams to integer streams. Combinatorial
functions are implicitly lifted to apply pointwise to their arguments: e.g., if
$\mathtt{x}=(x_i)_{i \in \Nat}$ and $\mathtt{y}=(y_i)_{i \in \Nat}$ are two
integer streams, then $(\mathtt{x+y}) = (x_i + y_i)_{i \in \Nat}$. Moreover, we
use a unitary delay, noted \Fby, such that $(\mathtt{x \Fby y})_i=x_0$ if $i=0$
and $y_{i-1}$ otherwise.

Such a program is classically compiled into a single function~$f$, which
computes the values of outputs and updates the system's state, from the values
of inputs and the current state. This function $f$ is then embedded inside a
periodic execution loop. Our contribution is to extend this classical
compilation scheme to a distributed framework: the result of the compilation of
a distributed system will consist of $n$ functions~$f_i$, one for each
location~$i$, which will compute the values of outputs, communication channels,
and local state, from the values of inputs, other incoming communication
channels, and the current local state.

\marc{Peu informatif. Que veux-tu dire?}
\gwen{Qu'il existe un schéma de compilation usuel des programmes
  synchrones, et ce qu'on veut obtenir comme schéma de compil. dans un
  cadre distribué. (paragraphe non modifié : c'est peu informatif
  parce que redondant avec une info déjà annoncée ? ou parce que c'est
  mal formulé ?)}
\marc{--> on comprend pas}
\gwen{Un peu modifié.}

\subsection{Language-based Distribution}
\label{sec:lang-based-distr}

We address \emph{functional} distribution, not achieved for the sake of
performance but because the system is intrinsically distributed. Distribution is
driven by the fact that some functions have a meaning only at some specific
locations and not at others. We can think, e.g., of a function returning the
value of a physical sensor and which has to be executed where the sensor is.
Therefore, locations will be defined by the functionalities they provide.

Designing such distributed systems is non-trivial, because of problems
such as the scheduling of communications or the type consistency of
the communicated data. The usual method, using architecture languages
like AADL~\cite{04:_aadl}, involves describing the system's
architecture by partitioning it in subsystems. Each subsystem can then
be defined separately, possibly with different languages. However, in
the case of tightly dependent subsystems, where conflicts between
architectural and functional modularity can occur, it is less
error-prone and more efficient to define the system as a whole,
together with architectural annotations. Our first contribution is to
provide language primitives to allow the programmer to describe the
architecture, and to express where some values are located, i.e., on
which location some computations are performed.

The architecture is described by the explicit declaration of the set
of existing locations and the links between them. At this point,
locations are symbolic: a location declaration introduces a symbolic
name, which will then be used to express the fact that a stream is
computed or available at this symbolic location. We define in
Section~\ref{sec:projection} a \emph{projection} operation which
produces, for each symbolic name, a single non-distributed synchronous
program to be executed at the physical location represented by this
symbolic name.

The syntax for declaring the physical location \texttt{A} is \texttt{loc~A}. The
existence of a communication link from \texttt{A} to \texttt{B} is declared by
\texttt{link~A~to~B}. Note that we distinguish \emph{communication links} from
\emph{communication channels}, introduced in Section~\ref{ssec:spatial-types-1}:
communication links, specified by the primitive \Link, state the \emph{ability}
to communicate from one location to another. In contrast, actual \emph{channels}
used by the distributed system are inferred by the type system.

\begingroup
\name{\f}{f}
\name{\g}{g}
\name{\h}{h}
\name{\A}{A}
\name{\B}{B}
\name{\c}{c}
\name{\x}{x}
\name{\y}{y}
\name{\z}{z}

The statement $e \At A$ means that every value used in the expression $e$
(streams and nodes composing its subterms) will be located at $A$. The
programmer does not need to express the localization of every value. Our second
contribution is to provide a type and effects
system~\cite{talpin92:_polym_type_region_effec_infer} whose double function is
to check the validity of the localization expressed w.r.t.\ the architecture,
and to infer the localization of non-explicitly located values. For instance,
the node \f given below consists of two computations \g and \h, respectively
located by the programmer on locations \A and \B, thanks to the \mbox{\texttt{at
    A}} and \mbox{\texttt{at B}} annotations.
\begin{code}
  \Letnode f(x) = z \With
        y = g(x) \At A
    \And z = h(y) \At B;
\end{code}
\endgroup
\vspace{-3mm}
Communications are abstracted, and thus not expressed by the programmer, neither
technically, nor concerning their place inside the code. The technical
expression of communications is left to the further phase of integration on
actual architecture: our method only deals with inferring the localization of
these communications, and their coherence throughout the distributed code. We
assume for now that communications can occur at any localization, and can
concern any value entirely concealed within a location (i.e., not the
distributed data structures, like distributed pairs). From a programmer's point
of view, this choice is a compromise between no control at all (communications
are possible everywhere) and absolute control (the programmer expresses every
communication).

\marc{Je ne comprends pas cet exemple. Ou sont les deux executables ?}
\gwen{Ajout de phrases d'explications aux paragraphes ci-dessus.}

\subsection{A Spatial Type System for Automatic Distribution}
\label{sec:spatial-type-system}


We place ourselves in a functional framework, where for the sake of modularity,
functions can neither be inlined nor analyzed dependently of their calling
context. We provide a special type system dedicated to the distributed execution
of the program, as an analysis provided to the programmer to help him ensure the
consistency of the distribution specification. We call it a \emph{spatial type
  system} and, when clear from context, we shall simply refer to it as a type
system. This spatial type system describes the localization of streams, and a
type-directed approach is followed to achieve code distribution. This also
allows us to preserve higher-order features, hence allowing the expression of
dynamic reconfiguration of nodes by application of other nodes as inputs.

The other motivation for using a type system is to achieve type
inference: in order not to force the programmer to specify everything
(i.e., the localization of each stream), spatial types will be
inferred from the available spatial annotations in the source. The
spatial type system also checks the consistency of these annotations
with the given architecture. Spatial consistency means, e.g., that
applying a node located on a location to a stream located elsewhere is
not correct. As we are in a functional context, spatial types will be
inferred for each defined node modularly.

\marc{pas d'emphase. Je ne comprends pas le paragraphe ci-dessus. Soit
  c'est super connu et une phrase suffit. Soit c'est pas connu et il
  faut l'expliquer precisement. Quelle est le papier qui raconte ca?
  Toujours mettre une reference si un point est connu.}
\gwen{Reformulation.}
%
%
\marc{enlever l'affaire des horloges dont on se moque ici. Dire
  simplement que, ``besides (...) typing)''.}
\gwen{Fait.}
A typed program is then automatically distributed by the compiler, by
extracting, for each declared location, one program strictly composed
of computations to be performed on this location, as well as added
communications from and to other locations in the form of added inputs
and outputs.

\marc{L'inference n'est pas tres clairement (et de maniere
  convaincante) amenee.  Le point est de dire que l'on veut d'une part
  decrire des contraintes d'architectures et introduire un mecanisme
  d'annotation dans le source. Ce mecanisme d'annotation doit etre
  coherent. Cette coherence est decrite par un ensemble de regles de
  typage.}
\gwen{J'ai essayé de faire mieux.}

The spatial type of a stream is the location where this stream is
located. In the case of a stream whose values are communicated via a
channel from one location to another, its spatial type is a set: it is
the set of locations where the stream will be available. The spatial
type of a node $f$ is written $t_i\ovfun{S}t_o$, where $t_i$ and $t_o$
are respectively the spatial types of $f$'s inputs and outputs, and
$S$ is the set of locations involved in the computation of $f$.  This
set of locations can be larger than the union of $t_i$ and $t_o$'s
sets of locations, since the computation of $f$ can involve
intermediary locations.

\subsection{Examples}
\label{sec:examples}

\begingroup
\name{\A}{A}
\name{\B}{B}
\name{\f}{f}
\name{\fu}{f1}
\name{\fd}{f2}
\name{\ft}{f3}
\name{\g}{g}
\name{\h}{h}
\name{\x}{x}
\name{\m}{m}

All the examples below assume the architecture declaration:
\begin{code}
  loc A;  loc B;  link A to B;
\end{code}
\vspace{-3mm}

The first example is a sequence of three nodes \fu, \fd and~\ft, each
assumed to be of spatial type
$\/\vs.c\At\vs\ovfun{\set{\vs}}c\At\vs$. \fu and \ft are localized by
the programmer, respectively on \A and \B. \fd is not explicitly
localized.
\begin{code}
  node g(x) = y3 with
        y1 = f1(x) at A
    and y2 = f2(y1)
    and y3 = f3(y2) at B
\end{code}
\vspace{-3mm}
\noindent This node will be given the spatial type $c\At A\ovfun{\set{A,B}}c\At
B$. As the localization of computations has to be done modularly, a spatial type
for \fd will be given once, among the two possibilities $c\At
A\ovfun{\set{A}}c\At A$ and $c\At B\ovfun{\set{B}}c\At B$. In contrast, since
there is no communication link from \B to \A, the following node will be
rejected by the type system:
\begin{code}
  node g'(x) = y3 with
        y1 = f1(x) at B
    and y2 = f2(y1)
    and y3 = f3(y2) at A
\end{code}
\vspace{-3mm}
Furthermore, it can be noted here that node \g cannot be used within a
located declaration. The following node will be rejected by our type
system:
\begin{code}
  node g'(m,x) = y with
    y = g(m,x) at A
\end{code}
\vspace{-3mm}



The second example involves a higher-order node: the node \h takes as input two
nodes \f and \g, and an input~\x, and applies \f to \x at one location, and then
\g to the result of the first application at another location. This example shows
also how a node can be defined with local locations for more modularity. These
new locations are introduced as a list between $[\ldots]$, can then be used
within the node. This higher-order node uses two location variables $\vs_1$ and
$\vs_2$:
\begin{code}
  node h [\(\vs\sb{1},\vs\sb{2}\)] (f,g,x) = z with
        y = f(x) at \(\vs\sb{1}\)
    and z = g(y) at \(\vs\sb{2}\)
\end{code}
\vspace{-3mm}
\texttt{h} receives then the spatial type:
\begin{multline*}
  \/\va,\vb,\vc.\/\vs_1,\vs_2:\set{\vs_1\comm\vs_2}.\\
\begin{array}{l@{}r@{}l}\relax
  \Bigl(& (\va\At \vs_1\ovfun{\set{\vs_1}}\vb\At \vs_1) & \\
  &\times(\vb\At \vs_2\ovfun{\set{\vs_2}}\vc\At \vs_2) & \\
  &\times(\va\At \vs_1) & \Bigr) \ovfun{\set{\vs_1,\vs_2}}\vc\At{\vs_2}
\end{array}
\end{multline*}
The set of constraints ($\set{\vs_1\comm\vs_2}$) is inferred from the
links required by the node. These constraints are resolved, with the
actual architecture, when this node is instantiated. A constraint
$\vs\comm\vs'$ is resolved, either by stating $\vs=\vs'=s$, or
with two locations $s$ and $s'$ such that there exists a communication
link from $s$ to $s'$ in the local architecture. Thus, the node
\texttt{h} can be instantiated in these two ways (assuming the
existence of two nodes \texttt{f} and \texttt{g}, both of spatial type
$\/\vs.c\At\vs\ovfun{\set{\vs}}c\At\vs$):
\begin{code}
      y1 = h (f at A,g at A,x1)
  and y2 = h (f at A,g at B,x2)
\end{code}
\vspace{-3mm}

We can observe than an arrow type appearing on the left of another arrow type
cannot comprise more than one location. This is caused by the form taken by the
distribution: since the projection operation on one node is performed on
locations, and not sets of locations, we cannot handle effect variables, unlike
other type and effect systems.
\endgroup

\subsection{Application}
\label{sec:appl-mult-softw}

As a concrete example, we consider the definition of a reception channel of a
software radio. A \emph{software radio}, or \emph{software-defined radio}, is a
radio in which components usually defined as hardware, e.g., demodulation or
filter components, are defined as software~\cite{mitola95}. This allows in
particular the reconfiguration, possibly dynamic, of these components.

Consider a reception channel composed of three main components: a pass-band
filter allowing the selection of the carrier wave, a demodulator component, and
a component allowing the analysis of the received signal, e.g., an
error-correction function. For the sake of performance, these components are
usually implemented on different architectural elements: the pass-band filter on
a FPGA, the demodulator on a digital signal processor (DSP), and the error
correction on a general-purpose processor (GPP).

Each component of this reception channel could easily be defined separately. But
in the case of software-defined radio, the system must support several
functionalities~\cite{jondral05:_softw_defin_radio}: each of these
functionalities must be engineered separately, and then integrated
together. Then, there is a conflict between distribution and functional
modularity issues.

\marc{Je ne vois pas l'aspect dynamicity dans le papier. Soit l'expliquer, soit
  l'enlever.}

\gwen{Ajout justif. ordre sup dans motivations de l'utilisation d'un système de
  types}

Let us study the case of a multichannel reception system that supports
the two mobile standards GSM and UMTS: the former involves a filter
for 1800 MHz frequencies, a GMSK demodulator, and a CRC /
convolutional error correction module, while the latter involves a 2
GHz filter, a QPSK demodulator, and a CRC / convolutional / turbo
codes error correction module.

\begin{figure}[H]
  \centering
  \resizebox*{\linewidth}{!}{\input{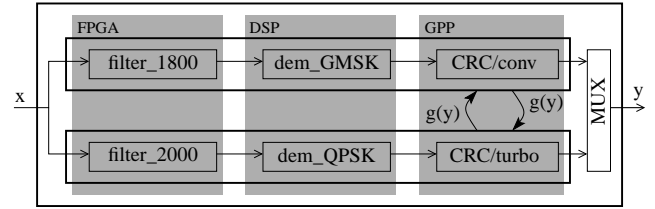}}
  \caption{Functional model of a multichannel software radio.}
  \label{fig:multichannel-sdr}
\end{figure}

Figure~\ref{fig:multichannel-sdr} shows an implementation of this
reception channel on a system composed of three hardware components: a
FPGA dedicated to the execution of the two pass-band filters, a DSP
for the demodulation functions, and a GPP for error-correction modules
and for the control of the whole system, i.e., in this case, the
switch between the two channels. This system has one input \texttt{x},
the radio signal from the antenna.  \texttt{y}~denotes the output
signal of the system, i.e., the decoded and corrected information
received by the channel. From this value, a function
\texttt{gsm\_or\_umts} (noted \texttt{g} on
Figure~\ref{fig:multichannel-sdr} for the sake of brevity), local at
GPP, computes what channel will be used at next instant. In this
figure, each location is graphically represented by a gray box.

In a classical context, designing this multichannel software radio would be
performed by separately programming each of the three hardware components, which
raises two problems. Firstly, there is no guarantee that the components interact
as specified: i.e., the 1800 filter with the GMSK demodulator, and so on. This
requires the MUX function to be duplicated on the three computing resources, so
as to guarantee the correctness of the system. This situation compromises the
modularity of the system. Secondly, each of the two channels corresponds to an
independent software entity.  Programming independently each hardware component
leads to the separate design, at least from some point of the design flow, of
closely related software components (e.g., filter and demodulator of the same
channel).

For the sake of modularity, this system would be better designed by considering
the channels independently, and not the hardware components. This situation
suggests adding primitives allowing to express the localization of streams
directly in the language. Such primitives should allow the programming of
software components independently of the architecture, handled as a separate
concern.  Thus, consistency analysis such as data typing could be performed on
the global program: communication channels could be typed and the data
consistency of the whole system could be checked. This way, inconsistencies due
to serialisation could be detected at compilation time.  \marc{Cette partie
  n'est pas claire.}

\begingroup
\name{\FPGA}{FPGA}
\name{\DSP}{DSP}
\name{\GPP}{GPP}

The code below implements this multichannel software radio, with our
extended language. The architecture consists of three locations,
\FPGA, \DSP, and \GPP, completely connected:
\begin{code}
\input{multichannel_sdr.ls}
\end{code}
\vspace{-4mm}
This implementation strictly follows the architecture of the system
described in Figure~\ref{fig:multichannel-sdr}. It shows the
declaration of three symbolic locations (\FPGA, \DSP, and \GPP).  We
assume that all filter, demodulation, and correction functions are
local ones, i.e., they are of spatial type
$\/\vs.c\At{\vs}\ovfun{\set{\vs}}c\At{\vs}$.  Since the conditional
construct comprises declarations that have to be executed on the set
of locations $\{\FPGA,\DSP,\GPP\}$, \texttt{c} is thus inferred to be
communicated to these locations. The conditional \If/\Then/\Else is
evaluated with the value of \texttt{c} at the previous instant. The
distribution of this example will put a copy of this \If/\Then/\Else
on these three locations. Finally, the expression \texttt{true fby c}
will be computed at \GPP, since the result of this expression has to
be communicated to the three locations where the conditional construct
will be duplicated.

By the same reasoning, we can infer that the spatial type of
\texttt{x} is \FPGA, and the one of \texttt{y} is \GPP. As a result,
the spatial type of the node \texttt{multichannel\_sdr} is:
\[(c\At\FPGA)\ovfun{\set{\FPGA,\DSP,\GPP}}(c\At\GPP)\]

\endgroup

\section{Formalization}
\label{sec:formalization}

We first define a synchronous dataflow core language
(Section~\ref{ssec:core-language-syntax}), and give its centralized semantics
(Section~\ref{ssec:centr-sem}) and its distributed semantics
(Section~\ref{ssec:distr-semant}). The centralized semantics is considered to be
the reference semantics and we only consider programs that react w.r.t.\ this
semantics. Programs that do not react (e.g., for typing or causality reasons)
are assumed to be rejected by other means~\cite{lucy:emsoft04}. The distributed
semantics allows us to give a meaning to location annotations. A spatial type
system is then presented (Section~\ref{ssec:spatial-types-1}). It is used to
both reject programs which cannot be distributed and to annotate every
expression from the source code with explicit locations.

\subsection{The Core Language Syntax}
\label{ssec:core-language-syntax}

\[
\begin{array}{l!{::=}l}
  P  & \cA \Pv d \Pv D\\
  \cA  &  \cA\Pv\cA \ou \Loc A \ou \Link A \To A\\
  s & \vs \ou A\\
  d  & \Letnode f[\ton{\vs}{,}](p) = e \With D \ou d \Pv d \\ 
  D  & p = e \ou p = x(e) \ou D \And D \ou \If e \Then D \Else D\\
  p  & p,p \ou x\\
  e  & i \ou x \ou (e,e) \ou \op(e,e) \ou e \Fby e \ou
  e\At s
\end{array}
\]

A program is made of an architecture description ($\cA$), a sequence
of node definitions ($d$) and a main set of equations ($D$).
An architecture description is a set of declarations of locations
($\Loc A$) or links ($\Link A\To A$) which state the existence of a
communication link from one location to another. A location $s$ is
either a location variable $\vs$, or a location constant $A$.
A node definition is composed of an expression and a set of
equations. A set of local locations $\set{\ton{\vs}{,}}$ can be
associated as location parameters of a node definition ($\Letnode
f[\ton{\vs}{,}](p)$ $= e\With D$).
Definitions $D$ define patterns of variables $p$, and are either single
equations ($p = e$), definitions naming the result of an application ($p =
x(e)$), parallel declarations ($D\And D$), or alternative declarations ($\If e
\Then D \Else D$).
An expression $e$ may be an immediate value ($i$), a variable~($x$), a pair
construction ($e,e$ ; pair destruction can be performed by pattern definitions),
a binary combinatorial operation ($\op(e,e)$, where $\op$ can be $(+)$,
$(-)$,\ldots), an initialized delay ($e \Fby e$), or an expression annotated
with an explicit location $s$ ($e\At s$).

\subsection{The Centralized Synchronous Semantics}
\label{ssec:centr-sem}

The purpose of the centralized semantics is to serve as a reference
semantics. This semantics does not take into account distribution
primitives. We first introduce auxiliary definitions. A value is
either an immediate constant ($i$), a pair or a function.
\[
\begin{array}{l}
  v ::= i \ou (v,v) \ou \x.e\With D\\
  R ::= [v_1/x_1,\ldots,v_n/x_n] \text{ s.t. } \/i\neq j, x_i\neq x_j\\
  S ::= R_1.R_2\ldots\\
\end{array}
\]
A reaction environment $R$ associates values to names and assumes that names are
pairwise distinct. $S$ denotes a sequence of reaction environments.

Given a sequence $d$ of node definitions $\Letnode
f_i[\vec{\vs}_i](x_i) = e_i\With D_i$, an initial global environment
$R_d$ is defined, holding $\lambda$-values of each $f_i$. This initial
environment will be given as input of the main program.
\[
  R_d = [\x_1.e_1\With D_1/f_1,\ldots,\x_n.e_n\With D_n/f_n] 
\]

The synchronous centralized semantics is defined by means of two
reaction predicates. \semcent{R}{e_1}{v}{e_2} states that in the
reaction environment $R$, the expression $e_1$ emits the value $v$ and
rewrites into the new expression~$e_2$. The predicate
\semcent{R}{D_1}{R'}{D_2} states that in the reaction environment $R$,
the declaration $D_1$ defines the reaction environment $R'$ and
rewrites into~$D_2$. The centralized execution of a program $P$ is
denoted \semprog{S_i}{P}{S_o}, meaning that under a sequence of
input environments $S_i = R_1.R_2\ldots$, the program $P = \cA\Pv
d\Pv D$ produces a sequence of output environments $S_o =
R'_1.R'_2\ldots$ such that:
\[
\infer
{
  \semcent{R_d,R_i,R_o}{D}{R_o}{D'}\\
  \semprog{S_i}{\cA\Pv d\Pv D'}{S_o}
}
{\semprog{R_i.S_i}{\cA\Pv d\Pv D}{R_o.S_o}}
\]

The rules for the reaction predicates are given in Figure~\ref{fig:cent-sem}.

\begin{figure}[H]
  \centering
  \begin{small}
    \begin{mathpar}
      \inferrule*[left=Imm]{}
      {\semcent{R}{i}{i}{i}}

      \inferrule*[left=Inst]{}
      {\semcent{R,[v/x]}{x}{v}{x}}

      \inferrule*[left=Fby]
      {
        \semcent{R}{e_1}{v_1}{e'_1}\\
        \semcent{R}{e_2}{v_2}{e'_2}
      }
      {\semcent{R}{e_1\Fby e_2}{v_1}{v_2\Fby e'_2}}

      \inferrule*[left=Op]
      {
        \semcent{R}{e_1}{i_1}{e'_1}\\
        \semcent{R}{e_2}{i_2}{e'_2}\\
        i = \op(i_1,i_2)
      }
      {\semcent{R}{\op(e_1,e_2)}{i}{\op(e'_1,e'_2)}}

      \inferrule*[left=Pair]
      {
        \semcentshort{R}{e_1}{v_1}{e'_1} \;\;\;\;
        \semcentshort{R}{e_2}{v_2}{e'_2}
      }
      {\semcent{R}{(e_1,e_2)}{(v_1,v_2)}{(e'_1,e'_2)}}

      \inferrule*[left=At]
      {\semcentshort{R}{e}{v'}{e'}}
      {\semcentshort{R}{e\At s}{v}{e'\At s}}



      \inferrule*[left=Def]
      {\semcent{R}{e}{(\ton{v}{,})}{e'}}
      {\semcent{R}{(\ton{x}{,}) = e}{[v_i/x_i]}{(\ton{x}{,}) = e'}}

      \inferrule*[left=App] 
      {
        R(f)=\p.e\With D \;\;\;\;\;
        \semcent{R}{p' = e \And p = e' \And D}{R'}{D'} 
      } 
      {\semcent{R}{p' = f(e')}{R'}{D'}}

      \inferrule*[left=And]
      {
        \semcent{R,R_2}{D_1}{R_1}{D'_1}\\
        \semcent{R,R_1}{D_2}{R_2}{D'_2}
      }
      {\semcent{R}{D_1\And D_2}{R_1,R_2}{D'_1\And D'_2}}

      \inferrule*[left=If-1]
      {
        \semcent{R}{e}{\True}{e'}\\
        \semcent{R}{D_1}{R'}{D_1'}
      }
      {
        \semcent{R}{\If e \Then D_1 \Else D_2}{R'}{\If e' \Then D_1' \Else D_2}
      }

      \inferrule*[left=If-2]
      {
        \semcent{R}{e}{\False}{e'}\\
        \semcent{R}{D_2}{R'}{D_2'}
      }
      {
        \semcent{R}{\If e \Then D_1 \Else D_2}{R'}{\If e' \Then D_1 \Else D_2'}
      }
    \end{mathpar}
  \end{small}
  \caption{Centralized synchronous semantics.}
  \label{fig:cent-sem}
\end{figure}

An immediate value emits itself and rewrites to itself (rule \rulename{Imm}). A
variable emits its current value as it is present in the reaction environment
(rule \rulename{Inst}). An initialized delay $e_1\Fby e_2$ emits the first value
of $e_1$, then the previous value of $e_2$ (rule \rulename{Fby}). An operation
is performed pointwisely on immediate values (rule \rulename{Op}). Pair
construction follow classical rules (rule \rulename{Pair}). Locations are not
taken into account here (rule \rulename{At}): annotations added by the
programmer do not alter the centralized semantics of the program (i.e., its
functionality). An equation $p = e$ emits the reaction environment defining the
variables contained in $p$ (rule \rulename{Def}). A sequential function
application is replaced by its body and argument definition (rule
\rulename{App}). Parallel equations are mutually recursive (rule
\rulename{And}). A conditional statement executes its first branch if its
condition is true (rule \rulename{If-1}) and its second branch otherwise (rule
\rulename{If-2}).

\subsection{The Distributed Synchronous Semantics}
\label{ssec:distr-semant}

The distributed semantics also operates on a program $P = \cA\Pv d\Pv D$, but
takes into account the architecture description and the explicit
locations. However, it remains a synchronous semantics in the sense that the
desynchronization due to the insertion of communications is not accounted
for. It defines a \emph{spatialized execution}: the values $\dv$ emitted by
expressions are now \emph{distributed values}, i.e., they are annotated with
location information stating how these values are distributed on the
architecture:
\[
\begin{array}{l!{::=}l}
  \dv & \lv \At A \ou (\dv,\dv) \ou \Lambda\vec\vs.\x.e\With D\\
  \lv & i \ou (\lv,\lv)\\
  \R & [\dv_1/x_1,\ldots,\dv_n/x_n] \text{ s.t. } \/i\neq j, x_i\neq x_j\\
  \S & \R_1.\R_2.\ldots\\
  G & \graph{\cS}{\cL}\\
\end{array}
\]

A distributed value $\dv$ is either a local value $\lv$, localized on the site
$A$ ($\lv\At A$), a distributed pair $(\dv,\dv)$, or a node. A sequence of
location parameters $\vec{\vs}$ is associated to nodes. A local value is either
an immediate value $i$, or a local pair $(\lv,\lv)$. $\R$ denotes a distributed
reaction environment, and $\S$ a sequence of distributed environments. $G$
denotes an architectural graphs composed of a set of locations $\cS$, and a set
of communcation links $\cL\subseteq\cS\times\cS$.

Several values can represent the same distributed value:
\begin{mathpar}
  (\lv_1, \lv_2) \At A = (\lv_1 \At A, \lv_2 \At A)\\
  
  \infdouble
  {\dv_1 = \dv_1'}{\dv_2 = \dv_2'}
  {(\dv_1, \dv_2) = (\dv_1', \dv_2')}
\end{mathpar}
These equalities mean that a local pair $(1,2)$, localized on the site $A$, can
indifferently be denoted by the distributed values $(1,2)\At A$ or $(1\At A,
2\At A)$. A distributed pair can be, for example, the pair $(1\At A, 2\At B)$ :
the first compound is located on $A$, and the second on $B$.




The operator $\loc(\cdot)$ gathers the set of locations from a distributed
value:
\[
\begin{array}{l!{=}l}
  \loc(i\At s) & \set{s}\\
  \loc((\dv_1,\dv_2) \At s) & \loc(\dv_1)\cup\loc(\dv_2)
\end{array}
\]

The operator $|\cdot|$ erases annotations from a distributed value to get a
centralized value, and extends straightforwardly to reaction environments:
\begin{mathpar}
  |i\At s| = i

  |(\dv_1,\dv_2)\At s| = (|\dv_1|,|\dv_2|)
\end{mathpar}

The distributed semantics is defined by means of two predicates refined from
their centralized versions. \semdist{\R}{\ell}{e_1}{\dv}{e_2} states that in the
distributed reaction environment $\R$, the expression $e_1$ emits the
distributed value $\dv$ and rewrites into $e_2$. $\ell$ represents the set of
locations involved in the computation of $\dv$. The predicate for declarations
is defined as well.
 
We denote by $\cS$ the set of declared constant locations, and by
$\cL\subseteq\cS\times\cS$ the set of declared communication links. The relation
$\cL$ defines the \emph{ability} to communicate, and not the actual existence of
communication channels, which will be inferred by the refined version of the
type system. $G$ denotes an architecture graph, composed of a set of locations
$\cS$, and a set of links $\cL$ between these locations.
An architecture description $\cA$ defines an architecture graph $G$:
the notation \defarch{G}{\cA}{G'} means that given the architecture
graph $G$, $\cA$ defines the new architecture graph $G'$. The rules
\rulename{Arch}, \rulename{Def-Loc} and \rulename{Def-Link} define
this predicate:
\begin{small}
\begin{mathpar}
  \inferrule*[left=Arch]
  {
    \defarch{\graph{\cS}{\cL}}{\cA_1}{\graph{\cS_1}{\cL_1}}\\
    \defarch{\graph{\cS_1}{\cL_1}}{\cA_2}{\graph{\cS_2}{\cL_2}}
  }
  {\defarch{\graph{\cS}{\cL}}{\cA_1 \Pv \cA_2}{\graph{\cS_2}{\cL_2}}}
  
  \inferrule*[left=Def-Loc]{}{\defarch{\graph{\cS}{\cL}}{\Loc
  A}{\graph{\cS\cup\set{A}}{\cL}}}
  
  \inferrule*[left=Def-Link]{A_1,A_2\in\cS}
  {\defhierarchy{\graph{\cS}{\cL}}{\Link A_1 \To A_2}{\graph{\cS}
      {\cL\cup\set{A_1\mapsto A_2}}}}
\end{mathpar}
\end{small}

For clarity, we assume that the architecture graph $G$ is global for subsequent
semantic rules. The annotated execution of a program $P$ is
\semdistprog{\hat{S}_i}{P}{\hat{S}_o}, meaning that under a sequence of input
environments $\hat{S}_i = \R_1.\R_2\ldots$, the program $P = \cA\Pv d\Pv D$
produces a sequence of output environments $\hat{S}_o = \R'_1.\R'_2\ldots$:
\[
\infer
{
  \defarch{\graph{\emptyset}{\emptyset}}{\cA}{G}\quad
  \semdist{\R_d,\R_i,\R_o}{\ell}{D}{\R_o}{D'}\quad
  \semdistprog{\hat{S}_i}{\cA\Pv d\Pv D'}{\hat{S}_o}
}
{\semdistprog{\R_i.\hat{S}_i}{\cA\Pv d\Pv D}{\R_o.\hat{S}_o}}
\]
where $\R_d$ is defined from the sequence of node definitions $d=\Letnode
f_i[\vec{\vs}_i](x_i)= e_i\With D_i$ as:
\[
  \R_d = [\Lambda\vec{\vs}_1.\x_1.e_1\With
  D_1/f_1,\ldots,\Lambda\vec{\vs}_n.\x_n.e_n\With D_n/f_n]
\]

The rules for the predicates \semdist{\R}{\ell}{e_1}{\dv}{e_2} and
\semdist{\R}{\ell}{D_1}{\R'}{D_2} are given in Figure~\ref{fig:dist-sem}. An
immediate value can be emitted anywhere (rule \rulename{Imm}). Rule
\rulename{Inst} defines the instantiation. A distributed value can be
communicated from location $s$ to location $s'$ if there exists a communication
link from $s$ to $s'$ (rule \rulename{Comm}). A binary operation can be
performed only on immediate values located on the same location $A$; the result
is located on $A$ as well (rule \rulename{Op}). An annotated expression must
involve at most the location stated for its computation (rule
\rulename{At}). An application involves choosing a set of constant locations,
and replacing location parameters by these locations in the expression and the
declaration (rule \rulename{App}). The other rules state that the computation of
a statement involves the union of the locations involved for the computation of
its compounds.




Lemma~\ref{lem:sem-dist-cent} states that if a program reacts with the
distributed semantics, then it reacts with the centralized one and
produces the same values.

\begin{lemma}
  \label{lem:sem-dist-cent}
  For all $D,D',\R,\R'$, if $\semdist{\R}{\ell}{D}{\R'}{D'}$, then
  there exists $R,R'$ such that $R = |\R|$, $R' = |\R'|$ and
  $\semcent{R}{D}{R'}{D'}$.
\end{lemma}


\subsection{Spatial Types}
\label{ssec:spatial-types-1}

For the sake of clarity, we first present a simplified version of the type
system. For projection, we refine this first version to take communication
channels into account (Section~\ref{sec:distribution}).

The syntax of spatial type expressions is:
\marc{mettre sigma avant. Rajouter t dans sigma aussi. Il y a un
  pb. avec la contrainte C dans sigma qui n'est pas utilisee dans la
  gen. C'est louche.}
\gwen{La généralisation oblige que $C$ soit vide (pour les
  fonctions). $C$ dans sigma n'est utilisé que pour $\Let x = e_1 \In
  e_2$. Explications ajoutées plus loin.}
\[
\begin{array}{r!{::=}l}
  \sigma & \relax\/\ton{\va}{,}.\/\ton{\vs}{,}:C.t\\
  t & t \ovfun{\ell} t \ou t \times t \ou tc \At s\\
  tc & c \ou \va \ou  tc \fun tc \ou tc \times tc\\
  \ell & \set{\ton{s}{,}}\\
  s & \vs \ou A\\
  H & H \At A \ou [x_1:\sigma_1,\ldots,x_n:\sigma_n]\\
  C & \set{s_1\comm s'_1,\ldots,s_n\comm s'_n}\\
\end{array}
\]
$H$ is the spatial typing environments. $H \At A$ denotes a located
environment, i.e., a typing environment from which every spatial type
will be forced to represent a value entirely located on~$A$.

\begin{figure}[H]
  \centering
  \begin{small}
    \begin{mathpar}
      \infer[Imm]{}
      {\semdist{\R}{\ell}{i}{i\At s}{i}}

      \inferrule*[left=Comm]
      {
        \semdist{\R}{\ell}{e}{dv\At s}{e'}\\
        (s,s')\in\cL
      }
      {\semdist{\R}{\ell\cup\set{s'}}{e}{dv[s'/s]\At s'}{e'}}

      \infer[Inst]
      {}
      {\semdist{\R,[\dv/x]}{\loc(\dv)}{x}{\dv}{x}}

      \inferrule*[left=Fby]
      {
        \semdist{\R}{\ell_1}{e_1}{\dv_1}{e'_1}\\
        \semdist{\R}{\ell_2}{e_2}{\dv_2}{e'_2}
      }
      {\semdist{\R}{\ell_1\cup\ell_2}{e_1\Fby e_2}{\dv_1}{|\dv_2|\Fby e'_2}}

      \infer[Op]
      {
        \semdist{\R}{\ell_1}{e_1}{i_1\At A}{e'_1}\\
        \semdist{\R}{\ell_2}{e_2}{i_2\At A}{e'_2}\\
        i = \op(i_1,i_2)
      }
      {\semdist{\R}{\ell_1\cup\ell_2}{\op(e_1,e_2)}{i\At A}{\op(e'_1,e'_2)}}

      \inferrule*[left=Pair]
      {
        \semdist{\R}{\ell_1}{e_1}{dv_1\At s_1}{e'_1}\\
        \semdist{\R}{\ell_2}{e_2}{dv_2\At s_2}{e'_2}
      }
      {\semdist{\R}{\ell_1\cup \ell_2}{(e_1,e_2)}
        {(dv_1\At s_1,dv_2\At s_2)\At s_1\sqcup s_2}{(e'_1,e'_2)}}\\



      \inferrule*[left=At]
      {\semdist{\R}{\set{s}}{e}{\dv}{e'}}
      {\semdist{\R}{\set{s}}{e\At s}{\dv}{e'\At s}}

      \inferrule*[left=Def] 
      {\semdist{\R}{\ell}{e}{(\ton{\dv}{,})}{e'}}
      {\semdist{\R}{\ell}{(\ton{x}{,}) = e}{[\dv_i/x_i]}{(\ton{x}{,}) = e'}}

      \inferrule*[left=App]
      {
        \R(f)=\Lambda\ton[1][n]{\vs}{,}.\p.e\With D\At s\\
        \set{\ton[1][n]{s}{,}}\subseteq \cS\\
        \semdist{\R}{\ell}
        {p' = e[\vec{s}/\vec{\vs}] \And p = e'\And D[\vec{s}/\vec{\vs}]}
        {\R'}{D'}
      }
      {\semdist{\R}{\ell}{p' = f(e')}{\R'}{D'}}\\

      \inferrule*[left=And]
      {
        \semdist{\R,\R_2}{\ell_1}{D_1}{\R_1}{D'_1}\\
        \semdist{\R,\R_1}{\ell_2}{D_2}{\R_2}{D'_2}
      }
      {\semdist{\R}{\ell_1\cup \ell_2}{D_1\And D_2}{\R_1,\R_2}{D'_1\And D'_2}}

      \inferrule*[left=If-1]
      {
        \semdist{\R}{\ell}{e}{\True\At s}{e'}\\
        \semdist{\R}{\ell'}{D_1}{\R'}{D_1'}\\
        \/s'\in\ell', s\comm s'
      }
      {
        \semdist{\R}{\ell\cup\ell'}{\If e \Then D_1 \Else D_2}{\R'}{\If e' \Then D_1' \Else D_2}
      }

      \inferrule*[left=If-2]
      {
        \semdist{\R}{\ell}{e}{\False\At s}{e'}\\
        \semdist{\R}{\ell'}{D_2}{\R'}{D_2'}\\
        \/s'\in\ell', s\comm s'
      }
      {
        \semdist{\R}{\ell\cup\ell'}{\If e \Then D_1 \Else D_2}{\R'}{\If e' \Then D_1 \Else D_2'}
      }
    \end{mathpar}
  \end{small}
  \caption{Distributed synchronous semantics.}
  \label{fig:dist-sem}
\end{figure}

We distinguish spatial type schemes~($\sigma$), which can be quantified, from
simple spatial types~($t$). A set of constraints $C$ can be associated to
quantification of location variables ($\/\ton{\vs}{,}:C.t$). We note
$\/\ton{\vs}{,}.t$ the scheme $\/\ton{\vs}{,}:\emptyset.t$. A simple spatial
type can be either a node type ($t \ovfun{\ell} t$), a pair type ($t \times t$),
or a located type ($tc\At s$). A located type can be either a stream type~($c$,
such as boolean, integer, etc.), a type variable~($\va$), a local function
($tc\fun tc$), or a local pair type ($tc\times tc$).  $\ell$ denotes sets of
locations. A location is either a location variable $\vs$, or a location $A$.

$C$ is a set of constraints between locations. A constraint $s_1\comm
s_2$ means that either $s_1 = s_2$, or there exists a communication
link from $s_1$ to $s_2$. Conversely, a declaration of communications
links $\cL$ leads to the set of constraints $\constraints(\cL)
=\set{s\comm s'|(s,s')\in\cL}$.

\marc{Je ne comprends pas l'ensemble des contraintes C. S'il n'y a
  qu'une seule variable dans un ensemble de contraintes, ca veut dire
  que l'on a des contraintes de la forme A < B, ce qui ne sert a rien
  (car soit c'est dans les defs et on peut l'enlever, soit ca n'y est
  pas et c'est faux), soit delta < A, soit A < delta. Ici, ca
  n'apporte pas grand chose, il suffit de faire une regle de
  sous-typage qui permet de remplacer les constantes par des
  variables, non?}
\gwen{Ce qui veut dire que $C$ n'apparaîtrait que dans $\sigma$, et
  qu'on a pas besoin de trimballer de contraintes dans toutes les
  règles. En fait le sys. de types tel qu'il est actuellement reflète
  ce qui est fait à l'implémentation : l'adéquation des contraintes
  avec l'architecture, et le fait d'affecter une valeur constante aux
  variables de sites, est fait à la généralisation. Mais je suis
  d'accord qu'il serait plus propre de les enlever.}
\gwen{Contraintes finalement déplacées.}

A value of spatial type $tc\At s$ is a value located on~$s$. A value of spatial
type $t_1 \ovfun{\ell} t_2$ is a node whose input is of spatial type~$t_1$,
whose output is of spatial type~$t_2$, and whose computation involves the set of
locations~$\ell$. The following equalities stand:
\[
\begin{array}{r!{=}l}
  (tc_1\times tc_2) \At s & (tc_1 \At s) \times (tc_2 \At s)\\
  (tc_1\fun tc_2)\At s & (tc_1 \At s) \ovfun{\set{s}} (tc_2 \At s)\\
\end{array}
\]
\marc{Il manque une regle de contexte. Dire sinon que on prend la plus
  petite relation verifiant les predicat precedents et fermee par
  contexte.}
\gwen{Pas sûr d'avoir compris. Les règles ci-dessous suffisent-elles ?}
\[
  \infdouble
  {t_1 = t_1'}{t_2 = t_2'}
  {(t_1\times t_2) = (t_1'\times t_2')}
  \qquad
  \infdouble
  {t_1 = t_1'}{t_2 = t_2'}
  {t_1\ovfun{\ell}t_2 = t_1'\ovfun{\ell}t_2'}
\]


The instantiation mechanism ensures the localization of a type
instantiated from a located environment:
\begin{multline*}
  (t[tc_1/\va_1,\ldots,tc_n/\va_n,s/\vs],C[s_1/\vs_1,\ldots,s_m/\vs_m])\\
  \leq\/\va_1\ldots\va_n\/\vs_1\ldots\vs_m:C.t
\end{multline*}
\[
  (tc \At s,C) \leq (H\At s)(x) \Leftrightarrow (tc \At s,C)\leq H(x)
\]

We note respectively $\FLV(t)$ and $\FTV(t)$ the set of free location
variables and free type variables of the type $t$. $\FLV$ and $\FTV$
are straightforwardly extended to typing environments.





A set of constraints $C$ is \emph{compatible} with a set of
communication links $\cL$, noted $\cL\models C$, iff $s\comm s'\in
C\land s\neq s'\Rightarrow (s,s')\in\cL$.

Before presenting our spatial type system, we introduce the following
notations:
\begin{itemize}
\item For a program $P$, the notation \defprogram{P}{t} means that the
  program $P$ is of spatial type~$t$.

\item For declarations (resp. expressions), the notation
  \spacetype{H}{G}{\ell}{D}{H'} (resp. \spacetype{H}{G}{\ell}{e}{t})
  means that, in the spatial type environment $H$ and the architecture
  graph $G$, the declaration $D$ (resp. the expression $e$) defines a
  new environment $H'$ (resp. is of spatial type $t$), and its
  computation involves the set of locations $\ell$.
\end{itemize}

The function $\locations(\cdot)$ gives the set of locations involved
in the spatial type given as argument. It is defined as:
\[
\begin{array}{ll}
  \locations(t_1\times t_2) & = \locations(t_1)\cup\locations(t_2)\\
  \locations(t_1\ovfun{\ell}t_2) & = \ell\\
  \locations(tc\At s) & = \set{s}
\end{array}
\]

The top-level declaration of a program is typed from the initial
environment~$H_0$, defined as:
\[
  H_0 = \left[
    \begin{array}[c]{l}
      \cdot\Fby\cdot : \/\va.\/\vs.\va\At\vs\times\va\At\vs\ovfun{\set{\vs}}\va\At\vs,\\
      \mathtt{(+)} : \/\vs.c\At\vs\times c\At\vs\ovfun{\set{\vs}}c\At{\vs},\ldots\\
    \end{array}
  \right]
\]

Our spatial type system is formally defined by the inference rules shown in
Figure~\ref{fig:typing}. Typing a program involves building an architecture
graph from the architecture description, and then using it to type the nodes and
the main declarations (rule \rulename{Prog}).

An immediate value can be used on any location (rule \rulename{Imm}).  Type
schemes can be instantiated (rule \rulename{Inst}). Typing a pair involves
stating that this pair has to be evaluated on the union of the sets of locations
on which each member of the pair has to be evaluated (rule \rulename{Pair}).
Typing a located expression ($e\At A$) involves building a located typing
environment (rule \rulename{At}). Communications are expressed as subtyping
(rule \rulename{Comm}).

The spatial type of a node consists of the spatial types of its inputs, the
computed expression, and the set of locations involved in this computation (rule
\rulename{Node}). The type of a node is generalized w.r.t.\ the set of locations
and links introduced by this architecture.

Typing an equation $x = e$ involves building a singleton typing environment
(rule \rulename{Def}).  Rule \rulename{App} states that an application must be
evaluated on the union of the set of locations where the node $f$ and its
argument $e$ must be evaluated, and the set of locations $\ell_1$ involved in
the computation of the node $f$.  Parallel declarations involve, for their
computations, the union of the sets and of locations involved in the computation
of their compounds (rule \rulename{And}).  Finally, typing an \If/\Then/\Else
declaration involves locating the condition expression on a location $s$, and
adding constraints that every location involved in declarations $D_1$ and $D_2$
must be accessible from $s$ (rule \rulename{If}).

\begin{figure}[H]
  \centering
  \begin{small}
    \begin{mathpar}
      \inferrule*[left=Prog]
      {
        \defarch{\graph{\emptyset}{\emptyset}}{\cA}{G} \;\;\;\;\;
        \spacetype{H_0}{G}{\ell}{d}{H} \;\;\;\;\;
        \spacetype{H,H_1}{G}{\ell'}{D}{H_1}
      }
      {\defprogram{\cA \Pv d\Pv D}{H_1}}\\

      \infer[Imm]{}
      {\spacetype{H}{G}{\set{s}}{i}{c\At s}}

      \inferrule*[left=Inst]
      {(t,C)\leq(H(x))\\\cL\models C}
      {\spacetype{H}{\graph{\cS}{\cL}}{\locations(t)}{x}{t}}

      \inferrule*[left=Pair]
      {
        \spacetype{H}{G}{\ell_1}{e_1}{t_1}\\
        \spacetype{H}{G}{\ell_2}{e_2}{t_2}
      }
      {\spacetype{H}{G}{\ell_1\cup\ell_2}{(e_1,e_2)}{t_1\times t_2}}\\

      \inferrule*[left=Def]
      {\spacetype{H}{G}{\ell}{e}{\ton{t}{\times}}}
      {\spacetype{H}{G}{\ell}{(\ton{x}{,}) = e}{[t_1/x_1,\ldots,t_n/x_n]}}

      \inferrule*[left=At]
      {
        \spacetype{H \At s}{\graph{\cS}{\cL}}{\ell}{e}{t}\\
        s\in\cS
      }
      {\spacetype{H}{\graph{\cS}{\cL}}{\ell}{e \At s}{t}}

      \inferrule*[left=Comm]
      {
        \spacetype{H}{\graph{\cS}{\cL}}{\ell}{e}{tc\TAt s}\\
        \cL\models s\comm s'
      }
      {
        \spacetype {H} {\graph{\cS}{\cL}} {\ell\cup\set{s'}} {e} {tc\TAt s'}
      }
      
      \infer[Node]
      {
        \spacetype{H,x_i:t_i,H_1}{\graph{\cS'}{\cL'}}{\ell_1}{D}{H_1}\\
        \spacetype{H,x_i:t_i,H_1}{\graph{\cS'}{\cL'}}{\ell_2}{e}{t}\\
        \cS' = \cS\cup\set{\ton[1][p]{\vs}{,}}\\
        \cL' \subseteq 
        \cL\cup(\set{\ton[1][p]{\vs}{,}}\times\cS)\cup(\cS\times\set{\ton[1][p]{\vs}{,}})\\
        \set{\ton[1][m]{\va}{,}} = \FTV(t)-\FTV(H)\\
        C = \constraints(\cL'\setminus\cL)\\
        \sigma = \/\ton[1][m]{\va}{,}.\/\ton[1][p]{\vs}{,}:C.
        (\ton{t}{\times})\ovfun{\ell_1\cup\ell_2}t
      }
      {\spacetype{H}{G}{\ell_1\cup\ell_2}
        {\Letnode f[\ton[1][p]{\vs}{,}](\ton{x}{,}) = e\With D}
        {[\sigma/f]}}\\

      \inferrule*[left=App]
      {
        \spacetype{H}{G}{\ell_2}{f}{t\ovfun{\ell_1}(\ton{t'}{\times})}\qquad
        \spacetype{H}{G}{\ell_3}{e}{t}
      }
      {\spacetype{H}{G}{\ell_1\cup\ell_2\cup\ell_3}{(\ton{x}{,}) = f(e)}
        {[t'_1/x_1,\ldots,t'_n/x_n]}}

      \inferrule*[left=And]
      {
        \spacetype{H}{G}{\ell_1}{D_1}{H_1}\\
        \spacetype{H}{G}{\ell_2}{D_2}{H_2}
      }
      {\spacetype{H}{G}{\ell_1\cup\ell_2}{D_1\And D_2}{H_1,H_2}}


      \inferrule*[left=If]
      {{\spacetype{H}{G}{\ell}{e}{c\At{s}}}\\
        {\spacetype{H}{G}{\ell_1}{D_1}{H'}}\\
        {\spacetype{H}{G}{\ell_2}{D_2}{H'}}\\
        {\cL\models\set{s\comm s'|s'\in\ell_1\cup\ell_2}}}
      {\spacetype{H}{G}{\ell\cup\ell_1\cup\ell_2}
        {\If e \Then D_1 \Else D_2} {H'}}
    \end{mathpar}
  \end{small}
  \caption{Spatial type system.}
  \label{fig:typing}
\end{figure}

We denote by $\dv:t$ the fact that the distributed value $\dv$ has
spatial type~$t$:
\begin{mathpar}
  \infer{}{\lv\At s:c\At s}
  
  \infer{\dv_1:t_1\\\dv_2:t_2}
  {(\dv_1,\dv_2):t_1\times t_2}
\end{mathpar}

We denote by $\R:H$ the type compatibility between $\R$ and $H$:
\begin{multline*}
  \R:H \Leftrightarrow \/x\in\dom(\R), x\in\dom(H)\\\land \exists(t,C)
  \text{ s.t. } (t,C)\leq H(x)\land \R(x):t
\end{multline*}

Theorem~\ref{thm:correction} states that if a program reacts with the
centralized semantics, and is accepted by our type system, then there exists a
spatialized execution such that the distributed values of this execution are
equal to the centralized ones. The types are preserved by this spatial
execution. The proof is omitted for lack of space.

\begin{theorem}[Soundness]
  \label{thm:correction}
  For all $D$,$D'$,$H$,$H'$,$R$,$R'$,$G$, if
  $\spacetype{H}{G}{\ell}{D}{H'}$ and $\semcent{R}{D}{R'}{D'}$, then
  there exists $\R,\R'$ such that $\semdist{\R}{\ell}{D}{\R'}{D'}$,
  $\R:H$, $\R':H'$, $|\R|=R$ and $|\R'|=R'$.
\end{theorem}








\marc{Pourquoi faire un deuxieme systeme de type ici vs ne pas
  seulement le faire au moment de la projection (dans la mesure ou
  celle-ci est dirigee par le typage? Ces points doivent etre
  discutes.)}
\gwen{Réponse partielle et non définitive : la projection sur les
  différents sites doit se faire à partir de la même ``coloration'',
  et des mêmes canaux inférés. C'est effectivement mal exprimé.}
\marc{Expliquer pourquoi on fait deux systemes.}
\gwen{Dit à la projection.}

\section{Distribution}
\label{sec:distribution}

\subsection{Principle}
\label{ssec:principle-dist}

Once programs have been typed, every expression is annotated with a
location that specifies where it has to be computed. Communications
are inserted when a value is produced at a location and used at
another. From this typed program, the compiler produces several new
programs --- one for every location $s$ --- erasing the code that is
not necessary at this location $s$. The run-time we have chosen is a
classical one for globally asynchronous locally synchronous (GALS)
systems: communications are done through FIFOs.

\begingroup
\name{\f}{f}
\name{\g}{g}
\name{\h}{h}
\name{\A}{A}
\name{\B}{B}
\name{\c}{c\_y}
\name{\x}{x}
\name{\y}{y}
\name{\z}{z}

We show below the result of the projection of the node \f of
Section~\ref{sec:lang-based-distr} on \A and \B, noted respectively
\texttt{f\_A} and \texttt{f\_B}. The distribution of this node will involve
adding a communication between these two computations. This communication will
take the form of an additional output on \texttt{f\_A} (named here \c, holding
the value \y computed on \A and used on~\B), together with an additional input
on \texttt{f\_B}. Original inputs and outputs are not suppressed: \abs denotes
an irrelevant value which will not be used on the current location. It is used
here to replace the output \z, whose computation is suppressed at~\A.

\begin{code}
  \Letnode f_A(x) = (\abs,c_y) \With
     c_y = g(x)

  \Letnode f_B(x,c_y) = z \With
     z = h(c_y)
\end{code}
\vspace{-2mm}
The semantically equivalent distributed system is then obtained by connecting
the input and output \c, holding the communicated value \y.  The program below
shows the distributed execution, using a FIFO materialized by
\texttt{send}/\texttt{receive} primitives, of the result of the projection of
the program \texttt{y = f(x)}.

\medskip
\begin{tabular}{p{0.4\linewidth}|p{0.4\linewidth}}
  \texttt{(y\_A,c\_y) = f\_A(x);}  & \texttt{receive(c\_y);}\\
  \texttt{send(c\_y)}              & \texttt{y\_B = f\_B(\abs,c\_y)}
\end{tabular}
\endgroup

\subsection{Example}
\label{sec:example-distrib}

The result of the projection of the two nodes of the
Section~\ref{sec:appl-mult-softw} on the location \texttt{DSP} is given on
Figure~\ref{fig:proj-dsp}. The projection of the \texttt{channel} node shows
that the node applications of \texttt{filter} and \texttt{crc} have been
removed, and that a new input \texttt{c1} (holding the value of \texttt{f}) and
a new output \texttt{c2} (holding the value of \texttt{d}) have been added. This
implies the addition, on the projection of the \texttt{multichannel\_sdr} node,
of two new inputs (\texttt{c2} and \texttt{c3}) and two new outputs (\texttt{c4}
and \texttt{c5}), one for each \texttt{channel} instance. The new input
\texttt{c1} of the projected \texttt{multichannel\_sdr} node holds the value
\texttt{c}.

\marc{Je ne comprends pas du tout le resultat de la figure 7. C'est quoi le main
  ? C'est grosso-modo le code de multichannel auquel on a ajoute des variables
  supplementaires. Il faut expliquer.}

\gwen{Paragraphe du dessus plus garni.}

\begin{figure}[H]
  \centering
  \setlength{\mynegesp}{-16mm}
\begin{code}
\input{multichannel_sdr.lsdsp}
\end{code}  
\vspace{-4mm}
  \caption{Result of the projection on \texttt{DSP}}
  \label{fig:proj-dsp}
\end{figure}

\subsection{Projection}
\label{sec:projection}

We will now define a type-directed operation of \emph{projection of an
  expression on a location $A$}. This operation is defined separately, as it has
to be performed on an already annotated program: links between values of each
projected program are defined by the channels inferred by the type system. The
projection will use a refined version of the type system, allowing the inference
of communication channels.

A channel is a location pair associated with a name, noted
\channel{A_1}{c}{A_2}: $c$ is the name of the channel, $A_1$ its source
location, and $A_2$ its destination location.  The set of channel names is
ordered by $<$, so as to keep consistency of inputs and outputs added, from the
node definition to node instances. $T$ denotes sequences of channels. The
concatenation of two sequences of channels, noted $T_1,T_2$, is defined iff
channel names in $T_1$ and $T_2$ are disjoint. We denote by $\epsilon$ the empty
sequence.

We note $\dom(T)$ the set of channel names of $T$. $T'\cong T$ means that the
sequences $T$ and $T'$ are equal, modulo channel renaming. This renaming allows
the multiple instanciations of node comprising communications.
\[
T'\cong T \Leftrightarrow T' = T[c'_1/c_1,\minildots,\!c'_n/c_n] \text{ where }
\set{c_1,\minildots,\!c_n} = \dom(T)
\]



The projection of a declaration $D$ on a location $A$ is noted
$\projspacetype{H}{G}{\ell/T}{D}{H'}{A}{D'}$, and results in a new
declaration~$D'$, containing only the computations to be performed on
$A$.  The projection of an expression $e$, of spatial type $t$, on a
location $A$, is noted $\projspacetype{H}{G}{\ell/T}{e}{t}{A}{e'/D}$,
and results in a new expression $e'$, as well as a declaration $D$,
containing channels outputs to be defined. A channel named $c$ in an
environment channel will be introduced as the variable $c$ as input
or output of the target program, $c$ assumed to be of different name
space than other variables of the source program.


We denote by $\epsilon$ the empty declaration, and by $\abs$ a value
which will never be used (i.e., void). For any declaration $D$,
$D\And\epsilon = \epsilon\And D = D$. For any expression $e$, we have
$(\abs\ e) = (e\ \abs) = \abs$.

Also, we note $T\uparrow A$ (resp. $T\downarrow A$) the set of
channels with origin (resp.\ destination) $A$:
\begin{mathpar}
  \begin{cases}
    \emptyset\uparrow A = \emptyset\\
    ([\channel{A_1}{c}{A_2}],T)\uparrow A =
    \begin{cases}
      [\channel{A_1}{c}{A_2}],(T\uparrow A) & \text{if } A_1 = A\\
      (T\uparrow A) & \text{else,}\\
    \end{cases}
  \end{cases}

  \begin{cases}
    \emptyset\downarrow A = \emptyset\\
    ([\channel{A_1}{c}{A_2}],T)\downarrow A =
    \begin{cases}
      [\channel{A_1}{c}{A}],(T\downarrow A) & \text{if } A_2 = A\\
      (T\downarrow A) & \text{else.}\\
    \end{cases}
  \end{cases}
\end{mathpar}

The projection rules are given in Figures~\ref{fig:proj-decl-1}
and~\ref{fig:proj-decl-2}.  Channels are used at communication points. If an
expression $e$ is sent from $A$ to $A'$ through the channel
$\channel{A}{c}{A'}$, then:
\begin{itemize}
\item for the projection on $A$, the communication involves sending a
  value: the resulting expression is void, and we add the definition
  of the channel $c$ as the result of the projection of $e$ on $A$
  (rule \rulename{Comm-P-From});

\item for the projection on $A'$, the communication involves receiving
  a value: the resulting expression is the channel holding this value
  (rule \rulename{Comm-P-To}).
\end{itemize}
Finally, if $A$ does not appear in the set of locations involved in
its computation, then the expression can be suppressed on $A$ (rule
\rulename{Suppr-P}).  Projections of a pair consist in the projection
of its compounds (rule \rulename{Pair-P}).

The projection of a located declaration and parallel declarations
involves the projection of its compound (rules \rulename{At-P} and
\rulename{And-P}).
%
%
The projections of applications and node definitions involve adding to
the inputs and outputs of the node, the channels used by this node
(rules \rulename{App-P} and \rulename{Node-P}). Nodes with local
architecture are assumed to be inlined. The relevance of the name
order appears here, as the order of the added inputs and outputs must
be consistent with every instances of these nodes, and for every
projection.

Projection of a conditional is divided in two rules: one for the
projection on a location where the conditional expression is computed:
this first rule shows the definition of every channel needed to send
this value to other locations where the conditional will be evaluated
(rule \rulename{If-P-From}); and one for the projection on a location
where the conditional expression has to be received: this expression
is then replaced by the name of the channel holding this value (rule
\rulename{If-P-To}).




\begin{figure}[H]
  \centering
  \begin{small}
    \begin{mathpar}
      \infer[Imm-P]{}
      {\projspacetype{H}{G}{\set{s}/\epsilon}{i}{c\At s}{A}{i/\epsilon}}

      \infer[Inst-P]
      {(t,C)\leq(H(x))\\\cL\models C}
      {\projspacetype{H}{G}{\ell/\epsilon}{x}{t}{A}{x_A/\epsilon}}

      \infer[At-P]
      {\projspacetype{H \At A}{G}{\ell/T}{D}{H'}{A}{D'}}
      {\projspacetype{H}{G}{\ell/T}{D \At A}{H'}{A}{D'}}

      \infer[Suppr-P]
      {A\not\in\ell}
      {\projspacetype{H}{G}{\ell/T}{e}{t}{A}{\abs/\epsilon}}

      \infer[Comm-P-From]
      {
        \projspacetype{H}{\graph{\cS}{\cL}}{\ell/T}{e}{tc\TAt A}{A}{e'/D}\\
        \cL\models A\comm s'
      }
      {
        \projspacetype
        {H}
        {\graph{\cS}{\cL}}
        {\ell\cup\set{s'}/T,[\channel{A}{n}{s'}]}
        {e}
        {tc\TAt s'}
        {A}
        {\abs/D\And c_n=e'}
      }

      \infer[Comm-P-To]
      {
        \projspacetype{H}{\graph{\cS}{\cL}}{\ell/T}{e}{tc\TAt s}{A'}{e'/D}\\
        \cL\models s\comm A'
      }
      {
        \projspacetype
        {H}
        {\graph{\cS}{\cL}}
        {\ell\cup\set{A'}/T,[\channel{s}{n}{A'}]}
        {e}
        {tc\TAt A'}
        {A'}
        {c_n/D}
      }

      \infer[Pair-P]
      {
        \projspacetype{H}{G}{\ell_1/T_1}{e_1}{t_1}{A}{e'_1/D_1}\;\;\;\;
        \projspacetype{H}{G}{\ell_2/T_2}{e_2}{t_2}{A}{e'_2/D_2}}
      {
        \projspacetype{H}{G}{\ell_1\cup\ell_2/T_1,T_2}
        {e_1,e_2}{t_1\times t_2}
        {A}
        {e'_1,e'_2/D_1\And D_2}
      }

      \inferrule*[left=Def-P]
      {\projspacetype{H}{G}{\ell/T}{e}{\ton{t}{\times}}{A}{e'/D}}
      {
        \begin{array}{c}
        \projspacetype{H}{G}{\ell/T\\}{(\ton{x}{,}) = e}{[t_1/x_1,\ldots,t_n/x_n]}
        {A}{(x_{1A},\ldots,x_{nA}) = e' \And D}
      \end{array}
      }

      \infer[App-P]
      {
        \projspacetype{H}{G}{\ell_2/T_2}{f}{t\ovfun{\ell_1/T_1}(\ton{t'}{\times})}{A}{f'/D_1}\\
        \projspacetype{H}{G}{\ell_3/T_3}{e}{t}{A}{e'/D_2}\\
        T'_1\cong T_1\\\\
        T'_1 \uparrow A = [\channel{A}{c_1}{A_1},\ldots,\channel{A}{c_m}{A_m}]\\
        T'_1 \downarrow A =
                          [\channel{A'_1}{c'_1}{A},\ldots,\channel{A'_p}{c'_p}{A}]
      }
      {
        \begin{array}{c}
          \projspacetype{H}{G}{\ell_1\cup\ell_2\cup\ell_3/T'_1,T_2,T_3\\}
          {(\ton{x}{,}) = f(e)}{[t_2/x]}
          {A}
          {
            (x_{1A},\ldots,x_{nA},c_1,\ldots,c_m) 
              = f'(e',c'_1,\ldots,c'_p)
            \And D_1 \And D_2
          }
        \end{array}
      }
      
      \infer[And-P]
      {
        \projspacetype{H}{G}{\ell_1/T_1}{D_1}{H_1}{A}{D_1'}\;\;\;\;\;\;
        \projspacetype{H}{G}{\ell_2/T_2}{D_2}{H_2}{A}{D_2'}
      }
      {
        \projspacetype{H}{G}{\ell_1\cup\ell_2/T_1,T_2}{D_1\And D_2}{H_1,H_2}
        {A}{D_1'\And D_2'}
      }


    \end{mathpar}
  \end{small}
  \caption{Rules for the projection operation (I).}
  \label{fig:proj-decl-1}
\end{figure}

The global meaning of a distributed program is then defined by the
parallelization of its projected declarations.

We note $\cS=\set{A_1,\ldots,A_n}$ the set of defined constant
locations where the source declarations are projected. The global
meaning of a declaration $D$, projected on the locations $\cS$, is
defined by:
\[
  D_1\And\ldots\And D_n
  \text{ where }
  \/i,\projspacetype{H}{G}{\ell/T}{D}{H'}{A_i}{D_i}
\]

In order to relate a target program with its source, we define a
relation on values, denoted \relval{\cdot}{\cdot}{\cdot}{\cdot}, such
that \relval{v'}{A}{t}{v} means that the value $v'$, emitted from an
expression of type $t$, represents the value $v$ at the location
$A$. We have:
\begin{mathpar}
  \infer
  {v' = v}
  {\relval{v'}{A}{t\At A}{v}}

  \infer
  {A \neq A'}
  {\relval{\abs}{A}{t\At A'}{v}}

  \infer
  {\relval{v'_1}{A}{t_1}{v_1}\\\relval{v'_2}{A}{t_2}{v_2}}
  {\relval{(v'_1,v'_2)}{A}{t_1\times t_2}{(v_1,v_2)}}
\end{mathpar}

We can then relate two reaction environments $R$ and $R_p$ w.r.t.\ a
typing environment $H$:
\[
  R \reaceq{H} R_p \text{ iff }
  \/x\in\dom(R),\/A\in\cS,\relval{R(x)}{A}{H(x)}{R_p(x_A)} 
\]

Theorem~\ref{thm:proj-decl} states that the projection operation is
correct, i.e., the projection of a source program $D$ into $D_i$ (for
every location $A_i$) defines a new target program $\ton{D}{\And}$,
which is semantically equivalent, taking into account spatial types'
values, with the source declaration $D$. The proof is omitted for lack
of space.

\begin{theorem}[Soundness of the declarations projection]
  \label{thm:proj-decl}
  For all $H$, $H'$, $D$, $D'$, $\ell$, $T$, $D_i$, $R$, $R'$, if
  $\semcent{R}{D}{R'}{D'}$, $\spacetype{H}{G}{\ell/T}{D}{H'}$ and
  $\forall i,\projspacetype{H}{G}{\ell/T}{D}{H'}{A_i}{D_i}$, then
  there exists $R_p,R_p',D_p,D_p'$ such that $D_p = D_1\And\ldots\And
  D_n$, $R\reaceq{H}R_p$, $\semproj{R_p}{D_p}{R_p'}{D_p'}$, and
  $R'\reaceq{H'}R_p'$.
\end{theorem}


\marc{Je me demande si ca ne serait pas plus clair de mettre l'exemple
  du debut apres chaque section technique pour illustrer ce qui se
  passe.}

\marc{Le systeme de type est assez clair et simple. Par contre, je
  n'ai pas reussi a comprendre le systeme faisant la projection.}

\begin{figure*}[tbp]
  \centering
  \begin{small}
    \begin{mathpar}
      \inferrule*[left=Node-P]
      {
        \projspacetype{H,x_i:t_i,H_1}{G}{\ell_1/T_1}{D}{H_1}{A}{D'}\\
        \projspacetype{H,x_i:t_i,H_1}{G}{\ell_2/T_2}{e}{t}{A}{e'/D_e}\\
        \set{\ton[1][m]{\va}{,}} = \FTV(t)-\FTV(H)\\
        \sigma = \/\ton[1][m]{\va}{,}.
            (\ton{t}{\times})\ovfun{\ell_1\cup\ell_2/T_1,T_2}t\\
        T_1,T_2\uparrow A = 
        [\channel{A}{c_1}{A_1},\ldots,\channel{A}{c_p}{A_p}]\\
        T_1,T_2\downarrow A = 
        [\channel{A'_1}{c'_1}{A},\ldots,\channel{A'_q}{c'_q}{A}]
      }
      {
        \begin{array}{c}
          \projspacetype{H}{G}{\ell_1\cup\ell_2/\emptyset\\}
          {\Letnode f(\ton{x}{,}) = e \With D}
          {[\sigma/f]}
          {A}
          {\Letnode f_A(x_{1A},\ldots,x_{nA},c'_1,\ldots,c'_q) 
            = (e',c_1,\ldots,c_p) \With D' \And D_e}
        \end{array}
      }

      \inferrule*[left=If-P-From]
      {
        \projspacetype{H}{G}{\ell/T}{e}{c\At{s}}{A}{e'/D}\\
        \projspacetype{H}{G}{\ell_1/T_1}{D_1}{H'}{A}{D_1'}\\
        \projspacetype{H}{G}{\ell_2/T_2}{D_2}{H'}{A}{D_2'}\\
        C=\set{s\comm s'|s'\in\ell_1\cup\ell_2}\\
        \cL\models C\\
        {T' = \channels(C)}\\
        {T' \uparrow A = [\channel{A}{c_1}{A_1},\ldots,\channel{A}{c_n}{A_n}]}\\
        {x\not\in\dom(H)}
      }
      {
        \begin{array}{c}
          \projspacetype{H}{G}{\ell\cup\ell_1\cup\ell_2/T,T_1,T_2,\channels(C)\\}
          {\If e \Then D_1 \Else D_2}
          {H'}
          {A}
          {x = e' \And c_1 = x \And \ldots \And c_n = x \And \If x \Then D_1' \Else D_2'}
        \end{array}
      }

      \inferrule*[left=If-P-To]
      {
        \projspacetype{H}{G}{\ell/T}{e}{c\At{s}}{A}{e'/D}\\
        \projspacetype{H}{G}{\ell_1/T_1}{D_1}{H'}{A}{D_1'}\\
        \projspacetype{H}{G}{\ell_2/T_2}{D_2}{H'}{A}{D_2'}\\
        C=\set{s\comm s'|s'\in\ell_1\cup\ell_2}\\
        \cL\models C\\
        {T' = \channels(C)}\\
        {T' \downarrow A = [\channel{A'}{c}{A}]}
      }
      {
        \projspacetype{H}{G}{\ell\cup\ell_1\cup\ell_2/T,T_1,T_2,\channels(C)}
        {\If e \Then D_1 \Else D_2}
        {H'}
        {A}
        {\If c \Then D_1' \Else D_2'}
      }
    \end{mathpar}
  \end{small}
  \caption{Rules for the projection operation (II).}
  \label{fig:proj-decl-2}
\end{figure*}

\section{Discussion}
\label{sec:discussion}

\subsection{Implementation}
\label{ssec:implementation}

The spatial type system presented in Section~\ref{ssec:spatial-types-1} relies
on a subtyping mechanism. It corresponds to the case where communications can
occur \emph{anywhere} in the code. This situation raises two problems. Firstly,
the implementation of type systems with subtyping mechanism is costly: usual
algorithms rely on the systematic application of the subtyping rule. Secondly,
this choice leads to a situation where the programmer has no control over where
the communications can occur. These problems, though orthogonal, can be
addressed together: giving some control to the programmer means restricting the
points where subtyping can be applied. We refine our type system in order to
address these two problems.

\marc{Je ne comprends pas ce paragraphe.}
\gwen{Un peu modifié. Devrait être plus clair avec les explications
  ajoutées dans les sections précédentes.}

We restrain communicated values to be variables introduced by equations ($x =
e$). Thus, in the program of Section~\ref{sec:lang-based-distr}, only \texttt{y}
and \texttt{z} can be communicated from one location to another. Then, we can
use a generalization mechanism to infer communication constraints, instead of
inferring them by subtyping. This corresponds to restricting the subtyping
mechanism to instantiation points.





This refined type system, as well as the projection operation presented in
Section~\ref{sec:distribution}, have been implemented in the Lucid Synchrone
compiler~\cite{lucid_synch_v3}. The independence of the method w.r.t. other
analysis allowed this implementation to be quite modular, implying very few
modifications within the rest of the compiler. The implementation consists to
generate distribution constraints, which are resolved modularly for each
node. We implemented a naive resolution algorithm, which can easily be replaced
by any existing distribution algorithm, taking into account efficiency issues
(for example, the AAA method of SynDex~\cite{sorel04:_syndex}).
This implementation has been tested on the software-defined radio example, which
is in its actual size composed of 150 lines of code, spread out into 25 Lucid
Synchrone nodes. We have tested our automatic distribution algorithm on a
benchmark suite composed of programs of few hundred of lines from the current
Lucid Synchrone release. This point is a good indication of the flexibility of
the extension proposed, allowing the distribution of programs not primarily
designed with distribution in mind. The scalability of this method is a direct
outcome of the use of a type system aiming at modular distribution; it has been
also applied on an ad hoc example composed of 10 Lucid Synchrone nodes,
comprising each 60 equations.

\subsection{Related Work}
\label{sec:related-work}


Various solutions have emerged in order to use synchronous languages for the
design of distributed systems. Some of them operate on a compiled model of the
program, by ``coloring'' atomic instructions with localization information,
inferred from input and output locations~\cite{dist-ieee}. The whole program is
first \emph{compiled} into an intermediate \emph{sequential} format, on which
the distribution is then applied. This format consists of a sequence of atomic
instructions, representing one computation step of new values carried on each
stream. Then, the distribution involves placing each instruction onto one or
several locations, taking into account the consistency of the control flow on
each location.  Another approach is to directly annotate the source program with
\emph{locations}, so as to define the localization of each variable of the
program: the distribution is then performed with regard to these annotations.
This approach has been applied to Signal~\cite{aubry96icse} as well as
Lustre~\cite{caspi03}. In both cases, the soundness of the distribution
algorithm has been proved~\cite{dist-jesa,maffeis-these}, meaning that the
combined behavior of the distributed fragments has the same functional and
temporal semantics as the initial centralized program. The originality of our
method resides in the fact that we use a spatial type system to check the
consistency of the distribution specifications inserted by the programmer, and
that we perform \emph{modular} distribution, allowing the expression of
higher-order features, applied for instance to dynamic reconfiguration of nodes
by application of other nodes as inputs. Since a Kahn semantics~\cite{Kahn-74}
can be given to our language, the semantic equivalence between the source
program and the synchronous product of the fragments resulting from the
projection is sufficient to describe an asynchronous distribution. While the
language presented has higher-order features, this method can easily be applied
to other language with comparable semantics such as Lustre. In contrast, more
general frameworks such as Signal cannot be addressed here for this reason.


Several approaches have been considered to solve the problem of data consistency
of distributed programs. A translation operation is presented
in~\cite{neubauer05:_from}, as well as an effect type system, in order to
automatically obtain a multi-tier application from an annotated source program.
Our proposition differs by the fact that our type system is not only a
specification, but also consists in what the authors called ``location
analysis'', thus allowing us to perform this analysis in a modular way, and on a
program comprising higher-order features. This last approach, as well as our
projection operation, can be compared with slicing
methods~\cite{ward07:_slicing}, as they consists in extracting specific parts of
a global program. Type systems have been used to ensure memory
consistency~\cite{talpin92:_polym_type_region_effec_infer}, or for pointer
analysis within a distributed architecture~\cite{liblit00:_type_dist_data}. The
\Acute language~\cite{sewell05:_acute} is an extension of \ocaml with typed
marshalling. Communication channels between two \Acute programs can also be
considered as typed, as the type of marshalled and unmarshalled data are
dynamically verified, at execution time. The consistency considered is between
separ\-ately-built programs, whereas our approach is to consider the programming
of a distributed system as one global program, allowing global static
verifications. Our approach can be compared with automatic partitioning: the
J-Orchestra system~\cite{liogkas04:_autom_partit} allows the user to assign
network sites to classes of Java programs. Then, this automatic partitioning
system transform the initial program into a distributed one, taking into account
distant or direct references. This last approach differs with our by the fact
that we integrate the distribution constructs within our language, which allows
relying distribution on semantical basis. Finally, \Oz and its distributed
extension~\cite{bal89:_progr_languag_distr_comput_system} proposes a way to
separate the functionality of a distributed application and its distribution
structure, by allowing the programmer to give a different distributed semantics
to every different object of a program. These two languages aims at loosely
coupled distributed systems without architecture constraints, whereas our
approach concerns strongly coupled architecture, and we aim to ensure the
consistency of the distribution w.r.t.\ one architecture, given the
communication constraints.

\subsection{Conclusion and Future Work}
\label{sec:conclusion}

We have proposed a spatial type system to solve the problem of automatic
distribution of dataflow programs. It is based on a core dataflow language,
which we have extended with distribution primitives to allow the programmer to
specify, on one hand his/her target distributed architecture, and on the other
hand where some nodes and/or variables are to be located.  The underlying
philosophy is the functional distribution, meaning that some functionalities of
the program must be computed at some precise location because they require some
specific sensors, actuators, and/or computing resources that are available only
at this location. In this context, we use type inference to decide at which
location each node must be computed, and at which points in the program
communication primitives must be inserted. The compilation of a correctly
spatially typed program produces one program for each computing location
specified by the programmer. We use abstract communication channels to exchange
a value between two locations and to synchronize them. Compiling each program
and linking with a dedicated library implementing those communication channels
then gives one binary code for each location. The refined version of the type
system, as well as the operation projection, has been implemented in a
synchronous dataflow compiler~\cite{lucid_synch_v3}.

Future work mainly involves allowing the description of more complex
architectures, with hierarchical locations or communication masking (e.g.,
MPSoCs): our proposal of architecture constraints is not sufficient to catch the
complexity of actual architecture of distributed embedded systems.  Yet, our
current constraints show the interest of a type system for checking the
consistency of a distributed program w.r.t.\ such constraints.

\marc{Other concern is the relation of this work with higher-order
  programming. We are interested now in using higher-order feature in a
  distributed framework.  Particularly, we want to consider the expression of
  dynamic reconfiguration of a resource by its dynamic load of its program from
  another resource, by code mobility.  }





\renewcommand{\/}{\corrital}




\end{document}